# Quad-cascade Picture of Electrokinetic Turbulence

Yanxia Shi [1,*], Jin'an Pang [1,*], Yueqiang Zhu [1,*], Ming Zeng [1], Keyi Nan [1], Yu Chen [1], Chen Zhang [1], Tianyun Zhao [2], Ce Zhang [1], Guangyin Jing [3,#], Kaige Wang [1,#], Jintao Bai [1], and Wei Zhao [1,#]

[1] State Key Laboratory of Photon-Technology in Western China Energy, International Collaborative Center on Photoelectric Technology and Nano Functional Materials, Institute of Photonics & Photon Technology, Northwest University, Xi'an 710127, China

[2] School of Automation, Northwestern Polytechnical University, Xi'an 710072, China

[3] School of Physics, Northwest University, Xi'an 710127, China

\# Correspondence: jing@nwu.edu.cn; wangkg@nwu.edu.cn; zwbayern@nwu.edu.cn

\* These authors contributed equally to this work;

Turbulence, ubiquitous in nature and across various systems, exhibits chaotic and intermittent fluctuations in space and time, defying precise prediction. For nearly a century, extensive efforts have been made to uncover the underlying universality and invariant laws from the immense disorder and chaotic nature of turbulence. While the celebrated Kolmogorov $-5/3$ law stands as a robust cornerstone, it falls short in capturing the diverse scaling behavior exhibited in turbulence influenced by external volume forces, like thermal convection and electrokinetic flows. This study proposes a general framework that couples the fluxes of kinetic energy and scalar variance, culminating in the formulation of a universal conservation law. This framework offers a comprehensive quad-cascade depiction of turbulence, enabling predictions that beyond the limitations of existing models. We illustrate this framework with microfluidic experiments on electrokinetic turbulence, wherein power spectra of concentration and velocity fluctuations exhibit the predicted scaling behaviors, providing remarkable agreement with theory. These findings not only deepen our understanding of the complete cascade process in turbulence driven by external volume forces but also hold promise for insights into other turbulent systems.

## 1. Introduction

Turbulence, a captivating phenomenon and a fundamental concept for engineers and theoretical scientists, has been the subject of study for over a century. Through various dynamic approaches, different types of turbulent phenomena, such as conventional hydrodynamic turbulence [1-4], turbulent thermal convection [5-7], electrohydrodynamic (or electrokinetic) [8,9], magnetohydrodynamic turbulence [10,11], quantum turbulence in Bose-Einstein condensate [12], and a new class of mesoscale turbulence in living matter [13], have been investigated to understand the energy cascade. In the case of hydrodynamic turbulence, the Richardson cascade theory proposes that large eddies break down into successively smaller ones until they are dissipated at the smallest scales. In the limit of negligible viscosity, a subrange is reached where a conserved kinetic energy flux ($\Pi_u$) is assumed, leading to the well-known Kolmogorov $-5/3$ law (K41 law) [14]. This scaling law has been extensively validated through both numerical simulations and experimental observations. Additionally, the Obukhov-Corrsin law (O-C law) [15,16], which includes the effects of passive scalar, has been established based on the assumption of constant fluxes of both kinetic energy and scalar variance ($\Pi_s$).

However, hydrodynamic turbulence is not the only mechanism capable of generating complex fluctuations. The situation becomes more intricate in turbulence driven by external volume forces, which strongly couple one or more scalar fields with the velocity field. In such cases, scalar turbulence arises, causing the transported substances (such as momentum, enthalpy, etc.) to experience fluctuations and randomness within turbulent flows. This, in turn, results in additional fluxes of scalar quantities. One prominent example is buoyancy-driven turbulence, such as stably stratified turbulence and turbulent thermal convection [17]. In these cases, Bolgiano-Obukhov law (BO59 law) [17-19], which is intrinsically based on a constant $\Pi_s$ but a non-constant $\Pi_u$, predicts scaling exponents of $-11/5$ and $-7/5$ for velocity and scalar spectra, respectively. These values deviate from the $-5/3$ scaling exponent of the K41 law and the additional scaling exponents in the O-C law. Interestingly, if both $\Pi_s$ and $\Pi_u$ are non-constant, a distinct scaling exponents coincide at a single value, i.e. $-3$, as proposed by Shur and Lumley [20]. Similarly, in electrokinetic turbulence, the Zhao-Wang model [21] also predicts a subrange where $\Pi_s$ remains constant while $\Pi_u$ varies.

Nevertheless, despite recent theoretical predictions [22,23], it is yet to be determined whether a subrange with a constant $\Pi_u$ but a non-constant $\Pi_s$ exists. More importantly, depending on the conservativeness, the fluxes of kinetic energy and scalar variance can be coupled into four combinations like a four-quadrant diagram. Therefore, it is intriguing to explore the possibility of unifying these well-known scaling laws into a comprehensive and universal framework by proposing a refined model. In this regard, we propose a quad-cascade picture of turbulence with a combined conservative equation as the coupled model. To experimentally validate this model, we specifically focus on electrokinetic (EK) turbulence in a microfluidics system, which provides a practical platform for investigating the turbulence under controlled conditions. This new model serves as a significant step towards unifying and comprehensively understanding the diverse aspects of turbulence driven by external volume forces.

## 2. Theory

EK turbulence is a style of turbulent flow attributed to the strong perturbation of electric body force (EBF), which can be generated when an electric field is applied on electric conductivity ($\sigma$) structures. It can be expressed by the control equations below [23,24]

$$\frac{D\boldsymbol{u}}{Dt} = -\frac{1}{\rho}\nabla p + \nu\nabla^2\boldsymbol{u} + \boldsymbol{N_1}\mathfrak{D}^{\frac{1}{4}}\sigma' \quad (1)$$

$$\frac{D\sigma'}{Dt} = -\boldsymbol{N_2}\cdot\boldsymbol{u} + D_\sigma\nabla^2\sigma' \quad (2)$$

$$\nabla\cdot\boldsymbol{u} = 0 \quad (3)$$

where $\boldsymbol{u}$ denotes velocity vector, $\sigma'$ is the fluctuations of electric conductivity, $\nu$ is kinematic viscosity, $D_\sigma$ is the effective diffusivity, $\mathfrak{D} = \Delta^2$ is biharmonic operator. $\boldsymbol{N_1}\mathfrak{D}^{1/4}\sigma'$ denotes EBF which relies on the spatial variation of $\sigma'$. Assume the external electric field is applied in y-direction ($\hat{y}$), say $E_y$, we have $\boldsymbol{N_1} = -\varepsilon E_y^2\hat{y}/\rho\langle\sigma\rangle$ which is the dimensional vector of EBF, and $\boldsymbol{N_2} = \nabla\langle\sigma\rangle$, with $\langle\sigma\rangle$ being the temporal averaging of electric conductivity. $\varepsilon$ is electric permittivity. Here, we assume $\nabla^2\sigma' \gg \nabla^2\langle\sigma\rangle$ and thus neglect $\nabla^2\langle\sigma\rangle$ term.

In Fourier space, we can define the modal turbulent kinetic energy and scalar variance as $E_u(\boldsymbol{k}) = \frac{1}{2}|\boldsymbol{u}(\boldsymbol{k})|^2$ and $E_\sigma(\boldsymbol{k}) = \frac{1}{2}|\sigma'(\boldsymbol{k})|^2$ respectively. According to previous reports [2,25,26], the transport equations of turbulent kinetic energy and scalar variance in Fourier space are

$$\frac{d}{dt}E_u(\boldsymbol{k}) = T_u(\boldsymbol{k}) - D_u(\boldsymbol{k}) + F_e(\boldsymbol{k}) \quad (4)$$

$$\frac{d}{dt}E_\sigma(\boldsymbol{k}) = T_\sigma(\boldsymbol{k}) - D_\sigma(\boldsymbol{k}) - F_\sigma(\boldsymbol{k}) \quad (5)$$

$$k_l u_l(\boldsymbol{k}) = 0 \quad (6)$$

with

$$T_u(\boldsymbol{k}) = \mathrm{Im}\left[\int k_l \widehat{u_l}(n)\widehat{u_q}(m)\widehat{u_q^*}(k)dk^3\right] \quad (7)$$

$$T_\sigma(\boldsymbol{k}) = \mathrm{Im}\left[\int k_l \widehat{u_l}(n)\widehat{\sigma'}(m)\widehat{\sigma'^*}(k)dn^3\right] \quad (8)$$

$$F_e(\boldsymbol{k}) = \mathrm{Re}\left[kN_{1_q}\widehat{\sigma'}(k)\widehat{u_q^*}(k)\right] \quad (9)$$

$$F_\sigma(\boldsymbol{k}) = \mathrm{Re}\left[N_{2_q}\widehat{\sigma'}(k)\widehat{u_q^*}(k)\right] \quad (10)$$

$$D_u(\boldsymbol{k}) = 2\nu k^2 E_u(\boldsymbol{k}) \quad (11)$$

$$D_\sigma(\boldsymbol{k}) = 2D_\sigma k^2 E_\sigma(\boldsymbol{k}) \quad (12)$$

where Re and Im represent the real and imaginary parts of the quantity. $\boldsymbol{k} = k_l\hat{x}_l$ and $k = |\boldsymbol{k}| = (k_l k_l)^{1/2}$, with $k_l$ being the wavenumber component in the $l^{\mathrm{th}}$ direction (denoted by $\hat{x}_l$). The wavenumbers have a relation of $\boldsymbol{k} = \boldsymbol{m} + \boldsymbol{n}$. $\widehat{u_q}$ is the Fourier transform of the $q^{\mathrm{th}}$ directional component of $\boldsymbol{u}$. $\widehat{\sigma'}$ is the Fourier transform of $\sigma'$. The asterisk represents a complex conjugate. $T_u(\boldsymbol{k})$ and $D_u(\boldsymbol{k})$ are the nonlinear turbulent kinetic energy transfer rate and dissipation rate, respectively. Similarly, $T_\sigma(\boldsymbol{k})$ and $D_\sigma(\boldsymbol{k})$ denote the nonlinear transfer rate of the scalar variance and scalar dissipation rate, respectively. Additionally, $F_e(\boldsymbol{k})$ denotes the energy feeding rate by the electric body force and $F_\sigma(\boldsymbol{k})$ signifies the scalar feeding rate of the electric conductivity.

For homogeneous and isotropic EK turbulence, considering statistical quasi-equilibrium state, according to Zhao's model [23], we have

$$\frac{d}{dk}\Pi_u(k) = F_e(k) - D_u(k) \quad (13)$$

$$\frac{d}{dk}\Pi_\sigma(k) = -F_\sigma(k) - D_\sigma(k) \quad (14)$$

where $\Pi_u$ and $\Pi_\sigma$ are the fluxes of kinetic energy and electric conductivity in $k$-space respectively, and

$$F_e(k) = \mathrm{Re}\left[kN_{1_q}\widehat{\sigma'}(k)\widehat{u_q^*}(k)\right] \quad (15)$$

$$F_\sigma(k) = \mathrm{Re}\left[N_{2_q}\widehat{\sigma'}(k)\widehat{u_q^*}(k)\right] \quad (16)$$

$$D_u(k) = 2\nu k^2 E_u(k) \quad (17)$$

$$D_\sigma(k) = 2D_\sigma k^2 E_\sigma(k) \quad (18)$$

where $E_u(k)$ and $E_\sigma(k)$ are the averaged power spectra of turbulent kinetic energy and the scalar variance over the spherical shell, $F_e(k)$ is the energy feeding rate by EBF due to electric conductivity field at wavenumber $k$, $F_\sigma(k)$ is the scalar feeding rate of electric conductivity at wavenumber $k$, $D_u(k)$ and $D_\sigma(k)$ are dissipation terms of kinetic energy and electric conductivity at wavenumber $k$ respectively.

When $\boldsymbol{N_2}$ is in y-direction, $\boldsymbol{N_1}$ and $\boldsymbol{N_2}$ are parallel to each other, from Eqs. (15) and (16), we get

$$F_e(k) - \frac{N_1}{N_2} F_\sigma(k) k = 0 \quad (19)$$

with $N_1 = |\mathbf{N_1}|$ and $N_2 = |\mathbf{N_2}|$. Then, in the subrange where EBF dominates the cascades of turbulent kinetic energy and scalar variance, the dissipation terms are neglected. From Eqs. (13), (14) and (19), we get

$$\frac{d}{dk}\Pi_u(k) + \frac{N_1}{N_2} k \frac{d}{dk}\Pi_\sigma(k) = 0 \quad (20)$$

Eq. (20) is the universal conservation equation in EK turbulence. It is established not only in the subrange where EBF is dominant, but also in the inertial subrange where $\Pi_u$ and $\Pi_\sigma$ are simultaneously constant. Thus, the $-5/3$ spectra of kinetic energy and scalar variance predicted in K41 and O-C law are unified into the model. Eq. (20) is finally transferred to the following equation

$$u_k + \frac{1}{N_1 N_2} k u_k^3 + \left(3k + \frac{5}{N_1 N_2} k^2 u_k^2\right) \frac{du_k}{dk} = 0 \quad (21)$$

according to

$$E_u(k) = u_k^2 / k \quad (22)$$
$$E_\sigma(k) = \sigma_k^2 / k \quad (23)$$
$$\Pi_u(k) = k u_k^3 \quad (24)$$
$$\Pi_\sigma(k) = k \sigma_k^2 u_k \quad (25)$$
$$u_k^2 = N_1 \sigma_k \quad (26)$$

where $u_k$ and $\sigma_k$ represent the velocity and electric conductivity components in $k$-space, respectively. Eq. (21) can be solved numerically with appropriate boundary conditions. With $u_k$ solved, $\sigma_k$, $E_u$ and $E_\sigma$ can be established according to Eqs. (26), (22) and (23) subsequently.

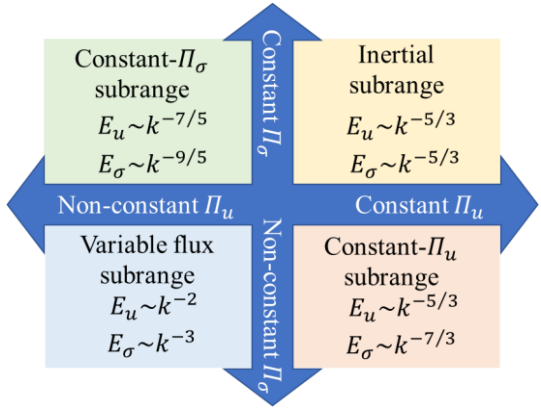

**Figure 1.** Quad-cascade picture of EK turbulence. [24]

As diagramed in Figure 1, four different solution sets can be predicted by solving Eq. (21), rendering quad-cascade picture in turbulence. All these subranges locate on the lower wavenumber side of the dissipation subrange. Therefore, ultrarich small-scale structures of velocity and electric conductivity can be predicted in EK turbulence. For inertial subrange and variable flux subrange, in the current investigation, they can emerge singly only. While for the constant-$\Pi_\sigma$ subrange and constant-$\Pi_u$ subrange, they can emerge simultaneously, as shown in Figure 2. The constant-$\Pi_\sigma$ subrange locates at larger wavenumbers, with $E_u \sim k^{-7/5}$ and $E_\sigma(k) \sim k^{-9/5}$. This is exactly what predicted in Zhao-Wang model [21]. The constant-$\Pi_u$ subrange is located at smaller wavenumbers, with $E_u \sim k^{-5/3}$ and $E_\sigma(k) \sim k^{-7/3}$. The existence of four solutions corresponding to the subranges can lead to significant difficulty to experimental observations. In the following sections, we will introduce the experimental observations of three of the four subranges.

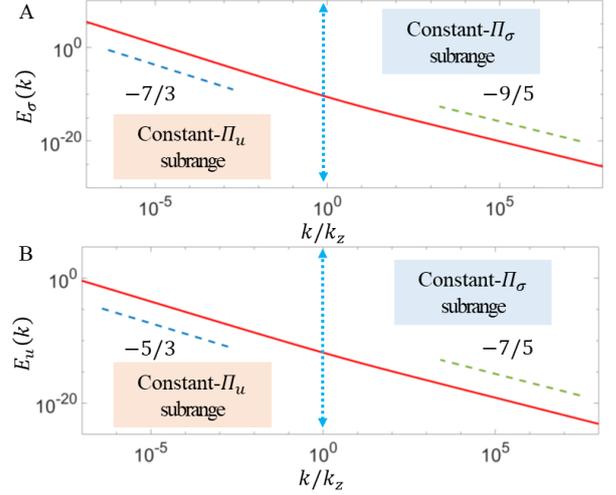

**Figure 2.** Theoretical results of $E_\sigma$ (A) and $E_u$ (B). Below $k/k_z = 1$ is constant-$\Pi_u$ subrange, while above $k/k_z = 1$ is constant-$\Pi_\sigma$ subrange.

## 3. Experimental setup and methods

To effectively generate EK turbulence, a sufficiently large electric Rayleigh number (as illustrated in Eq. 29 below) is required. This can be realized with a strong electric field which is more achievable in a microchamber. Thus, the experiments are conduct in a microfluidic chip. The structure of electric conductivity field in EK turbulence is investigated by flow visualization via laser induced fluorescence. The structure of velocity field is characterized by a state-of-art laser induced fluorescence photobleaching anemometer (LIFPA). [27-29]

### 3.1. Electrokinetic micromixer

The micromixer used in this experiment was fabricated using a standard UV lithography, as schematically shown in Figure 3. The upper layer of the micromixer is a Y-shaped microchannel made by polydimethylsiloxane (PDMS), and the lower layer is a cover glass with a thickness of 0.15 mm. The micromixer is 100 μm high ($h$), 9 mm long ($l$), with an initial width ($w_0$) of 620 μm at the entrance. Two cylindrical platinum electrodes with a diameter of 1 mm are symmetrically placed on both sides of the splitter plate in the micromixer, acting as the side walls of the expansion channel and provide an electric field. A splitter plate with a trailing edge at the entrance was fabricated to maintain a sharp interface of the two solutions delivered into the electric field.

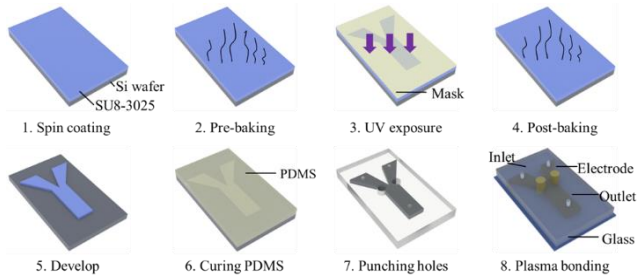

**Figure 3.** Schematic diagram of the fabrication of the micromixer. Two platinum rods (1 mm diameter) serving as electrodes and side walls as well are assembled at the inlet of mixing chamber.

In the experiments, one solution is de-ionized (DI) water that has lower electric conductivity ($\sigma_1$) and the other is prepared by phosphate buffered saline (PBS) (HyClone, SH30256.01, USA) solution with higher electric conductivity ($\sigma_2$), as shown in Figure 4A. They have an electric conductivity ratio of $\sigma_1:\sigma_2 = 1:5000$. Each of the two solutions has 3 μL/min flow rate. The Reynolds number of the bulk flow is $Re = Uw_0/\nu = 1$. To generate EK turbulence, an AC electric field of 130 kHz was applied by an arbitrary function generator (Tektronix, AFG3102C, USA) and a high voltage amplifier (Trek, PZD700A, USA). The actual output voltage was monitored by an oscilloscope (Tektronix, MSO 2022B, USA).

### 3.2. Flow visualization

The scalar EK turbulence in a micromixer was studied by flow visualization through laser induced fluorescence in an inverted epi-fluorescence microscopy system. For this purpose, solution 2 was prepared with a fluorescence dye, fluorescein sodium salt (Sigma, 46970-100G, Germany, $C_{20}H_{10}Na_2O_5$, molecular weight 376.27), which has an excitation peak wavelength at 460 nm and emission peak wavelength at 515 nm. The concentration is 500 μmol/L. To maintain the electric conductivity ratio between the deionized water and fluorescent solution at 1:5000, PBS has been applied as well. Since the fluorescence dye concentration is proportional to the concentration of ions in the solutions, the distribution of fluorescence intensity can directly reflect that of electric conductivity.

The schematic of the system is shown in Figure 4A. It consists of a 473 nm continuous-wave laser (MW-BL-473/200mW, CNI, China), a SCMOS camera (Edge 4.2LT, PCO, Germany), an inverted fluorescence microscope (NIB900, Nexcope, China),
an objective lens (10X NA0.32, Leica, Germany), a dichroism mirror and other optical elements. The laser passes through the microscope, reflected on the dichroic mirror, and irradiates the microchannel. The fluorescent solution emits fluorescent signals which transmit the dichroic mirror and a bandpass filter, then, are captured by the SCMOS camera. The dichroic mirror (Di01-R488/543/635, Semrock, USA) has high transmittance at 500 nm ~ 528 nm and high reflectivity at 380 nm ~ 491 nm. The fluorescent images of mixing were captured with 2048×2048 pixels, under an exposure time of 1 ms at a 40.58 fps frame rate. The spatial resolution of the system is around 900 nm.

### 3.3. Laser induced fluorescence photobleaching anemometer (LIFPA)

The velocity fluctuations of EK turbulence were measured by a LIFPA system [27,28,30], which enables flow velocity measurement by detecting the fluorescence intensity of fluid after photobleaching under laser illumination. In LIFPA measurement, both solutions are prepared with a fluorescence dye, Coumarin 102 (Sigma Aldrich, USA, $C_{16}H_{17}NO_2$, molecular weight 255.32), with the same concentration at 100 μmol/L. The fluorescence dye has an excitation peak wavelength at 390 nm and emission peak wavelength at 470 nm. The electric conductivity ratio of the two solutions is carefully maintained as 1:5000.

LIFPA is developed on a home-developed confocal microscope, as shown in Figure 4B. The system uses a continuous wave laser (MDL-III-405-500, CNI, China) with a wavelength of 405 nm as illumination source. The power of the laser is 11 mW. An acousto-optic modulator (1206C-2-1002, Isomet, USA) is used to modulate the beam temporally. Since only the first-order diffraction spot from acousto-optic modulator is used as the excitation beam, which has a poor beam quality, we applied a spatial light filter (SFB-16DM, OptoSigma, USA) to improve the beam quality. The light beam is collimated by a lens and subsequently reflected by a dichroic mirror (Di03-R405/532-t1-25×36, Semrock, USA), guided into the objective lens (PlanApo 100× NA1.4 oil immersion, Olympus, Japan). The fluorescence signal passes through the objective lens, the dichroic mirror, and two OD4 bandpass filters (470/100 nm and 470/10 nm) in turn. Then, the fluorescence signal is collected by a multimode fiber (25 μm in diameter, 400-550 nm band, M67L01, Thorlabs, USA) and detected by a single photon counter (H7421, Hamamatsu, Japan). Our previous researches have demonstrated the LIFPA system has a spatial resolution of 180 nm and temporal resolution of O(10 μs).[31] For a long term

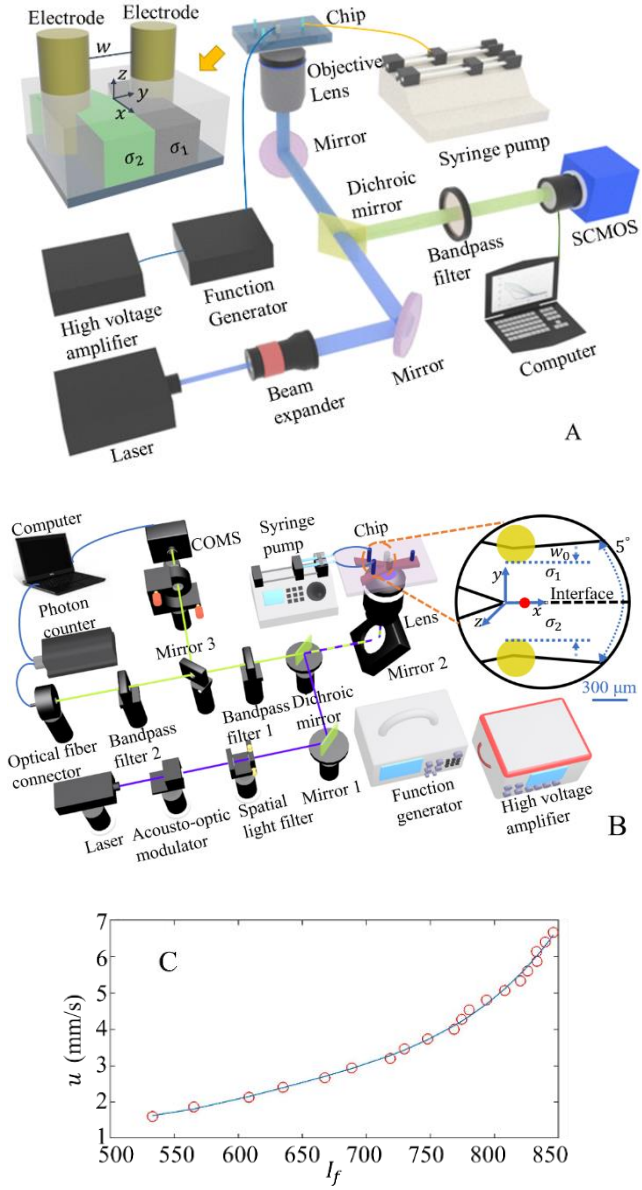

**Figure 4.** Experimental systems. (A) Flow visualization system through laser induced fluorescence in the investigation of EK turbulence, including an inverted fluorescent microscope, EK micromixer and electric system. (B) LIFPA system for EK turbulence, includes a home-developed confocal microscopy system, the EK micromixer and electric system. (C) A typical velocity calibration curve in LIFPA velocity measurement.

measurement, the sampling rate is 1 kHz only.

Since LIFPA measures flow velocity through measuring fluorescence intensity, a velocity calibration curve is acquired to establish the relationship between flow velocity and fluorescence intensity, as shown in Figure 4C. The velocity calibration curve is nonlinearly fitted by a fifth order polynomial. During velocity measurement, the fluorescence intensity ($I_f$) detected from the LIPFA system is transferred into flow velocity by the polynomial. Thus, the velocity time trace can be obtained for further statistics analysis.

### 3.4. Correction of the illumination field

The fluorescence intensity distribution in flow visualization strictly relies on the light intensity distribution of illumination. Although many efforts have been made in the development of microscopy systems, the influence of nonuniform illumination is inevitable, especially in widefield imaging. To overcome the spatial variation of the fluorescence intensity due to nonuniform illumination, a correction on the fluorescence intensity field ($I(i,j)$) has been applied before data analysis. The correction is carried out on the basis of dark ($I_{dark}(i,j)$, Figure 5A) and white ($I_{white}(i,j)$, Figure 5B) fields as

$$I_{corr}(i,j) = [I(i,j) - I_{dark}(i,j)]M(i,j) \quad (27)$$

with the correction matrix being

$$M(i,j) = \frac{\max[I_{white}(i,j) - I_{dark}(i,j)]}{I_{white}(i,j) - I_{dark}(i,j)} \quad (28)$$

where $I_{corr}(i,j)$ is the fluorescence intensity field after correction. $\max[I_{white}(i,j) - I_{dark}(i,j)]$ is the maximum value of $I_{white}(i,j) - I_{dark}(i,j)$ for all $i$ and $j$. The dark field (Figure 5A) represents the background noise level of the imaging system and is irrelevant to the illumination. In contrast, the white field (Figure 5B) is captured when the microchamber is filled with uniform fluorescence dye, resulting in its intensity distribution that is proportional to the light field of illumination. The influence of image correction can be clearly observed in Figure 5C-F for both instantaneous fluorescence intensity field and the mean field.

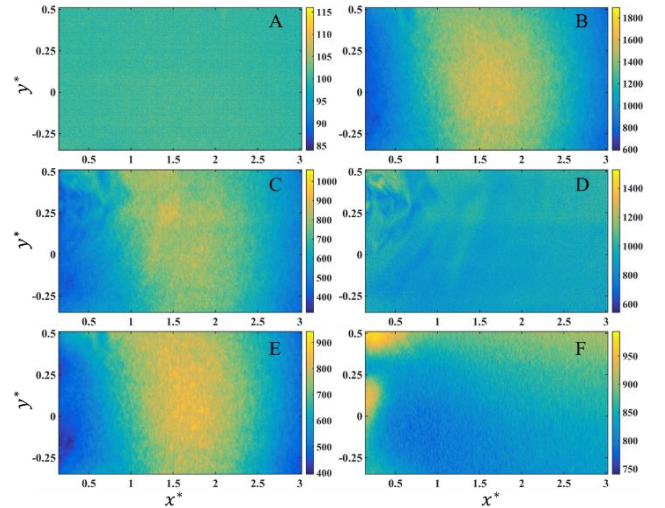

**Figure 5.** Light intensity distributions. (A) Dark field and (B) white field. (C) Instantaneous fluorescence intensity field before correction. (D) The corrected version of (C). (E) Mean field of fluorescence intensity before correction. (F) The corrected version of (E). $x^* = x/w_0, y^* = y/w_0$.

## 4. Results

### 4.1. Flow visualization

Figure 6 illustrates the progression towards EK turbulence. Without an electric field, the flow remains laminar and stratified (Figure 6A), characterized a distinct fluid interface at the center of the microchannel, showing the two streams with a large electric conductivity ratio. The fluctuations caused by external vibrations are negligibly small. Mixing in this case is primarily driven by molecular diffusion.

When an AC electric field is applied, the EBF can be sufficiently strong to overcome the influence of viscosity. This can be evaluated by an electric Rayleigh number[32] as

$$Ra_e = 4\varepsilon E_w^2 w_0^2 (1-\beta^2)(\sigma_2 - \sigma_1)/\rho \nu D_\sigma (\sigma_2 + \sigma_1) \quad (29)$$

where $E_w = V/w_0$ is the applied electric field intensity, $V$ is the applied peak-to-peak voltage, $\beta = 4\pi f_f \varepsilon/(\sigma_1 + \sigma_2)$ is a dimensionless frequency with $f_f$ being AC frequency. The higher the $Ra_e$, the stronger the EBF relative to viscosity. Therefore, the inertia of fluids can be sufficiently strong to realize turbulence.

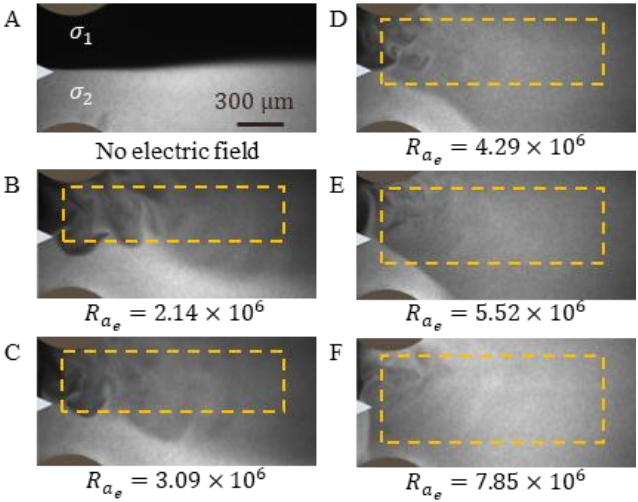

**Figure 6.** Flow visualization by laser induced fluorescence to show the mixing process of the two streams. The brown zones represent part of the electrodes and the dashed boxes indicate the calculation region of statistical quantities and scalar spectra. Six different AC electric fields of different voltages, e.g. 0 Vp-p (A), 67.2 Vp-p (B), 80.8 Vp-p (C), 95.2 Vp-p (D), 108 Vp-p (E) and 128.8 Vp-p (F), are applied at 130 kHz. The corresponding $Ra_e$ increases from 0 to $7.85 \times 10^6$.

When $Ra_e = 2.14 \times 10^6$, counter-rotating vortex patterns emerge under the influence of EBF. The interface between the low and high electric conductivity streams inclined towards the low electric conductivity region, with a spreading angle of 122°.

As $Ra_e$ increases beyond $3.09 \times 10^6$, the mixing becomes more pronounced, extending to the higher electric conductivity region, resulting in a spreading angle towards 180°. The entire mixing chamber becomes characterized by extensive mixing, allowing for finer scalar structures. At $Ra_e = 7.85 \times 10^6$, as inferred from the fluorescence intensity, uniform electric conductivity throughout the cross-section of the mixing chamber is achieved within a short distance ($x^* = 0.2$) from the trailing edge. The scalar field is approximately homogeneous in both large and small scales, as can be inferred from the mean fluorescence intensity field $\langle I \rangle$, the fluorescence intensity fluctuations $\langle I'^2 \rangle$ and their dissipation $\varepsilon_C = D_C \langle \nabla I' \cdot \nabla I' \rangle$ ($D_C$ is the molecular diffusivity of fluorescence dye) in Figure 7 respectively.

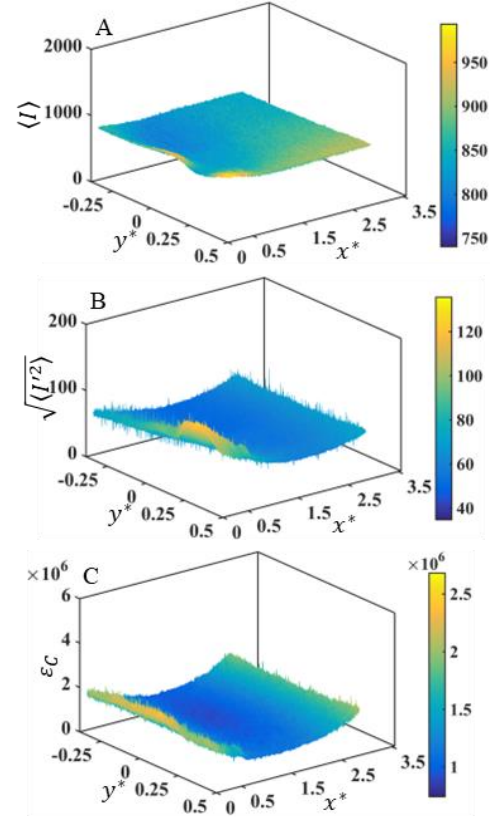

**Figure 7.** Statistics of the concentration field at $Ra_e = 7.85 \times 10^6$ calculated in the dashed box of Figure 6(F). (A) Mean fluorescence intensity field $\langle I \rangle$. (B) Fluorescence intensity fluctuations evaluated by $\langle I'^2 \rangle$. (C) Dissipation field $\varepsilon_C$.

### 4.2. Scalar spectra

The transport of scalar variance along wavenumber can be elucidated by the spatial spectra ($E_\sigma$) of electric conductivity, based on the fluorescence intensity fluctuation $I'$ [33]. Without electric field, $E_\sigma$ is flat with small magnitudes attributed to noises which do not affect the distinguishment of the spectral components in higher wavenumber region.

At the position $y^* = -0.12$ (note, $E_\sigma$ is computed in a region as exemplified in the dashed box of Figure 6B. The position $y^*$ indicates the lower bound of the region), at a lower $Ra_e = 2.45 \times 10^6$, a wide subrange with $-7/3$ slope (by nonlinear fitting) can be observed from $E_\sigma$ first (Figure 8A). This is more observable from its compensated spectrum $E_C = k^{-\alpha} E_\sigma$ (Figure 8B), with $\alpha$ being the slope of the scaling subrange. As $Ra_e$ is further increased to $3.27 \times 10^6$, the $-7/3$ spectrum shifts towards lower wavenumbers, accompanied by the appearance of a $-9/5$ spectrum at higher wavenumbers. Eventually, the $-9/5$ spectrum replaces the $-7/3$ spectrum at $Ra_e = 4.29 \times 10^6$, indicating a gradual evolution of cascade processes under increasing control parameter. As one of the important findings of this investigation, the experimental observations support the theorical predictions shown in Figure 2A from two aspects.

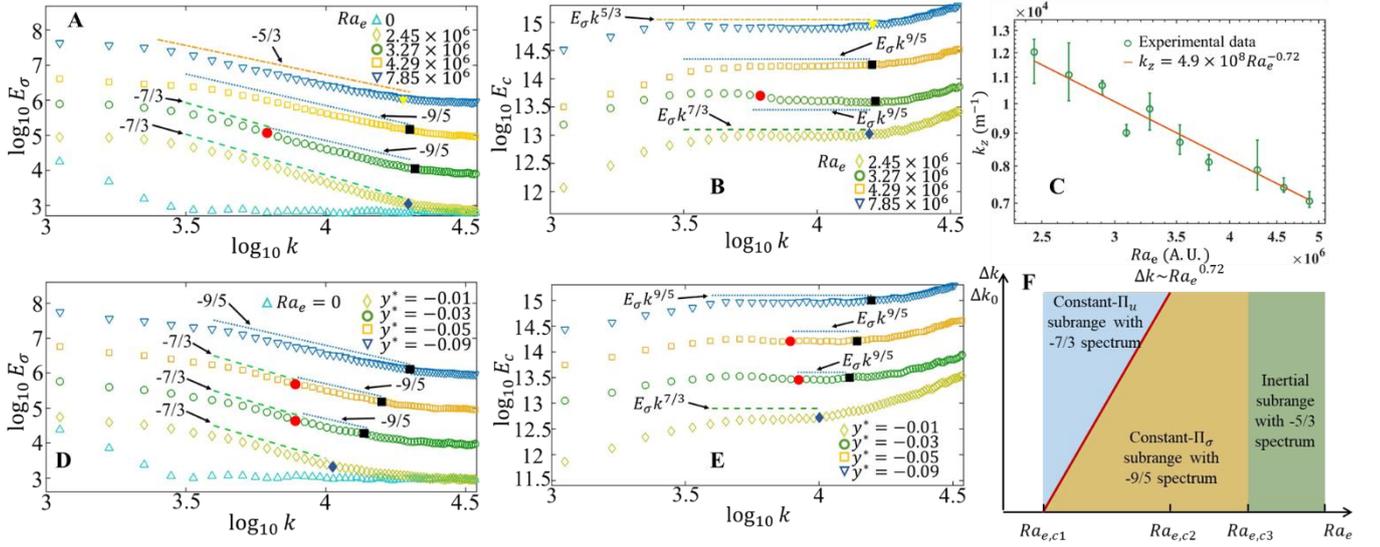

**Figure 8.** Concentration spectra $E_\sigma$ vary with $Ra_e$ and spanwise positions. Here, $k = 1/l$, with $l$ being length scale. (A) $E_\sigma$ vs $Ra_e$ at $y^* = -0.12$ (i.e. in the region of $y^* \geq -0.12$). From the top to bottom, $E_\sigma$ is vertically shifted by $10^3$, $10^2$, $10^1$, $10^0$ and $10^0$ respectively, for better clarity. The red dots indicate the intersection points of the $-7/3$ and $-9/5$ spectra. The blue diamond represents where the $-7/3$ spectrum terminates. The black squares denote the termination points of the $-9/5$ spectrum. The yellow triangle represents where the O-C $-5/3$ spectra terminate. (B) Compensated scalar spectra $E_C$ vs $Ra_e$ at $y^* = -0.12$. From the top to bottom, $E_C$ is vertically shifted by $10^{4.8}$, $10^{3.4}$, $10^{2.8}$, and $10^0$ respectively. (C) $k_z$ vs $Ra_e$ in both experiments and numerical simulations. (D) $E_\sigma$ vs $y^*$ at $Ra_e = 4.58 \times 10^6$. From the top to bottom, $E_\sigma$ is vertically shifted by $10^3$, $10^2$, $10^1$, $10^0$ and $10^0$ respectively. (E) Compensated scalar spectra $E_C$ vs $y^*$ at $Ra_e = 4.58 \times 10^6$. From the top to bottom, $E_\sigma$ is vertically shifted by $10^{4.2}$, $10^{3.5}$, $10^{2.8}$, and $10^0$ respectively. (F) Diagram of the width ($\Delta k$) of the scaling subranges in $E_\sigma$ with $Ra_e$. $\Delta k$ is determined by both the geometrical scale of the turbulent system and the performance of measurement techniques. $\Delta k_0$ is the maximum width. $Ra_{e,c1}$, $Ra_{e,c2}$ and $Ra_{e,c3}$ represent three critical $Ra_e$. At $Ra_{e,c1}$, the $-7/3$ spectrum emerges. At $Ra_{e,c2}$, the $-9/5$ spectrum completely replace $-7/3$ spectrum. And at $Ra_{e,c3}$, the inertial subrange with $-5/3$ spectra emerge.

On one hand, both the $-9/5$ and $-7/3$ slopes predicted from the numerical computations have been observed in $E_\sigma$ by experiments (Figure 8A). Both Zhao-Wang model [21,34] and the quad-cascade model has demonstrated that $-9/5$ spectrum is a constant-$\Pi_\sigma$ subrange, where $\Pi_\sigma$ remains quasi-constant, while $\Pi_u \sim k^{2/5}$ is nonconstant. In contrast, the $-7/3$ spectrum represents a constant-$\Pi_u$ subrange, where $\Pi_u$ remains quasi-constant, but $\Pi_\sigma \sim k^{-2/3}$. $\Pi_\sigma$ decreases rapidly with wavenumber accompanied by a decreasing transport capability until it reaches saturation in the constant-$\Pi_\sigma$ subrange at higher wavenumber.

On the other hand, the experimental results clearly demonstrated that the $-9/5$ spectrum is located in the higher wavenumber region of the $-7/3$ spectrum. This is also what we predict theoretically from the numerical computations (Figure 2A).

When $Ra_e$ is further increased to a sufficiently large value, e.g. $7.85 \times 10^6$, both the $-7/3$ and $-9/5$ spectra are absent, replaced by the well-known O-C $-5/3$ spectrum. This observation suggests the generation of an inertial subrange of scalar cascade within this microscale configuration. The direct observation of the O-C spectrum in a microchannel flow supports the findings of Wang et al. [35], who conducted temporal measurements on concentration fluctuations using single-point laser-induced fluorescence via a confocal microscope. Unfortunately, the evolution of $E_\sigma$ from $-9/5$ spectrum to $-5/3$ spectrum was neither experimentally nor theoretically observed in this investigation.

## 4.3. Characteristic wavenumber between constant-$\Pi_u$ and constant-$\Pi_\sigma$ subranges

The characteristic wavenumber ($k_Z$, red dot in Figure 8A) that connects the $-7/3$ and $-9/5$ spectra is further experimentally studied. $k_Z$ is a function of $N_1 N_2$ and $u_{k,1}$ which is the boundary value of $u_k$ at lower wavenumbers. According to Zhao's model [23], their relationship can be expressed as $k_Z \sim (N_1 N_2)^a u_{k,1}^b$, with $a \approx 3$ and $b = -6$. Since $N_1 N_2 \sim Ra_e$ and $u_{k,1} \sim Ra_e^g$ ($g$ to be determined by experiments), we approximately have $k_Z \sim Ra_e^{3-6g}$. The experimental results indicate $k_Z \sim Ra_e^{-0.72}$ (Figure 8C). The corresponding $g = 0.62$.

Therefore, the evolution of scaling subranges in EK turbulence, which is the second important finding of this investigation, can be inferred as shown in Figure 8F. When $Ra_e$ exceeds a critical value ($Ra_{e,c1}$) where the width of the $-7/3$ spectrum reaches maximum, the constant-$\Pi_u$ subrange becomes shrink with $Ra_e$. In the meanwhile, the constant-$\Pi_\sigma$ subrange becomes broadened, with the width of the $-9/5$ spectrum increasing as $Ra_e^{0.72}$ approximately. When $Ra_e$ further increases beyond another critical value ($Ra_{e,c2}$), the $-9/5$ spectrum completely replaces the $-7/3$ spectrum. Further increasing $Ra_e$ to over $Ra_{e,c3}$, the constant-$\Pi_\sigma$ subrange can transit to inertial subrange where $E_\sigma$ exhibits $-5/3$ spectrum. Beyond cutoff wavenumbers (marked in Figure 8A) of these scaling subranges, there should be a scalar dissipation subrange with an exponential decay along $k$. However, due to the relatively high noise level, the exponential decay was not observed. Therefore, the influence of $Ra_e$ on the cutoff wavenumbers and the scalar dissipation subrange cannot be evaluated from the current investigation. Additionally, the variable flux subrange was not observed as well.

## 4.4. Locality of solutions

There could a series of reasons may affect the observation of all the four subranges and their interrelationship. In the numerical computations, even if with the same $u_{k,1}$, we can still obtain all the four solutions simultaneously from Equation (21) (for details, please refer to [23]). Before figuring out the underlying conditions for each solution, we cannot simply exclude the possibility of coexistence of the solutions and the corresponding subranges in the flow field. Therefore, even though the scalar field shows approximate homogeneity, the structures of electric conductivity and the corresponding spectra are studied on different spanwise positions, as shown in Figure 8D and E.

It can be seen, as the measurement position moves from the center ($y^* = -0.01$) towards the sidewall ($y^* = -0.09$), the $-7/3$ spectrum appears initially and gradually transits to the $-9/5$ spectrum. Accordingly, the width of $-7/3$ spectrum and $k_Z$ both increase with $y^*$. Conversely, the width of $-9/5$ spectrum and $k_Z$ both decrease with $y^*$. The evolution of $-7/3$ spectrum and $-9/5$ spectrum along spanwise direction clearly indicates the scaling subrange of EK turbulence are local and simultaneously coexisted. This is the third important finding of this investigation.

## 4.5. Velocity spectrum and structure function

EK turbulence is generated according to the flow disturbance driven by EBF, which is inherently determined by scalar field, i.e. electric conductivity in this investigation. In the quad-cascade picture of EK turbulence (Figure 1), the four subranges can be distinguishable from $E_\sigma$, rather than $E_u$, since the inertial subrange and constant-$\Pi_u$ subrange share the same scaling indices. Therefore, unlike the commonly studied passive scalar turbulence where velocity field is dominant, electric conductivity field is more representative in EK turbulence. However, a direct measurement on velocity fluctuation is also necessary to comprehensively understand the structure of EK turbulence.

In this investigation, the velocity fluctuation is measured by LIFPA method, as introduced above. Since LIFPA is a single-point detection method, only the temporal variation of velocity can be obtained, as shown in Figure 9A. It can be seen under the influence of EBF, the flow is highly violated and random.

The spatial spectrum of velocity is calculated from the temporal variation of velocity, relying on Taylor's frozen hypothesis [36]. When $Ra_e = 2.89 \times 10^6$, according to Figure 8A, a constant-$\Pi_u$ subrange with $E_\sigma \sim k^{-7/3}$ should be observed. The corresponding velocity spectrum should be $E_u \sim k^{-5/3}$ (Figure 2B). This is consistent to the experimental result in Figure 9B. Similarly, when $Ra_e = 4.52 \times 10^6$, a constant-$\Pi_\sigma$ subrange with $E_\sigma \sim k^{-9/5}$ should be present as inferred from Figure 8A. The velocity spectrum in the subrange should be $E_u \sim k^{-7/5}$, which is also observed in experiments (Figure 9C). The experimental $E_u$ again supports the model of quad-cascade process of EK turbulence.

We also examined the energy flux through third-order structure function of velocity fluctuations, i.e. $S_u^3(l) = \langle |[u'(x+l) - u'(x)]^3|\rangle$. $S_u^3(l)$ is tightly related to $\Pi_u$. If $S_u^3(l) \sim l^{\xi_3}$, the corresponding $\Pi_u \sim k^{1-\xi_3}$.

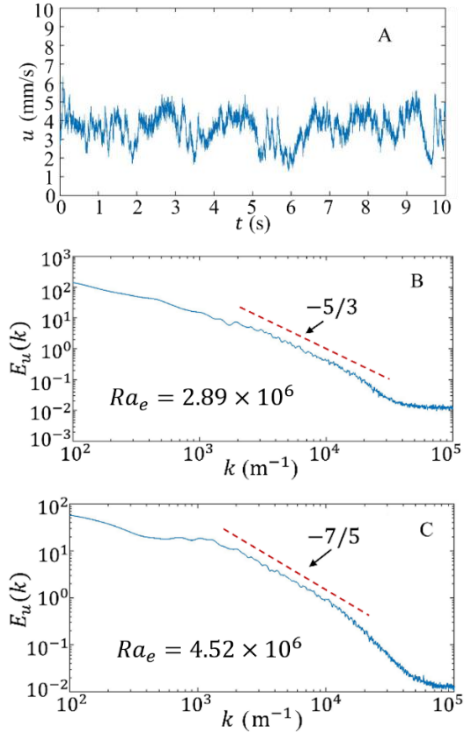

**Figure 9.** Velocity time trace and spectra. (A) typical time trace of velocity measured by LIFPA, where $Ra_e = 2.89 \times 10^6$. The sampling rate is 1 kHz. (B) Experimental results of $E_u$ at $Ra_e = 2.89 \times 10^6$. (C) Experimental results of $E_u$ at $Ra_e = 4.52 \times 10^6$.

When $Ra_e = 2.89 \times 10^6$, a constant-$\Pi_u$ subrange is observed, as supported by both $E_\sigma$ (Figure 8A) and $E_u$ (Figure 9B). In this subrange, theoretically we have $\Pi_u \sim k^0$. Thus, $\xi_3 = 1$. This is exactly what we can see from Figure 10A. When $Ra_e = 4.52 \times 10^6$, a constant-$\Pi_\sigma$ subrange is observed with $\Pi_u \sim k^{2/5}$. The corresponding $\xi_3 = 3/5$ is also experimentally validated in Figure 10B. Thus, the features of the constant-$\Pi_u$ subrange and constant-$\Pi_\sigma$ subrange are further supported through energy flux by experiments.

## 5. Discussions

To this end, we have demonstrated the existence of quad-cascade processes in EK turbulence from different perspectives. As proposed in this model, it is crucial to couple the fluxes of kinetic energy with the scalar variance to establish the conservation law. The EK turbulence is not a special case shows the strict coupling. Buoyancy-driven turbulence, as discussed by Verma and his collaborators [22,37], also shows similar properties. The strict coupling between the vector field (e.g., velocity) and scalar field naturally gives rise to four subranges that exhibit distinct scaling laws in different flow systems, as illustrated in Figure 11. Thus, the quad-cascade model provides a comprehensive framework for understanding the generalized scaling laws applicable to various turbulent systems.

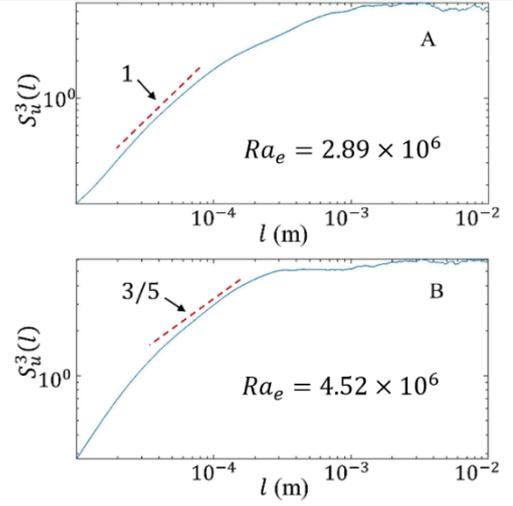

**Figure 10.** Velocity structure function. (A) $S_u^3$ vs $l$ at $Ra_e = 2.89 \times 10^6$. (B) $S_u^3$ vs $l$ at $Ra_e = 4.52 \times 10^6$. the data lengths are over $4 \times 10^5$.

On one hand, for instance, the quad-cascade model can guide the investigations in buoyancy-driven turbulence, EBF-driven turbulence (e.g. electrohydrodynamic/electrokinetic turbulence), magnetohydrodynamic turbulence and etc, particularly on numerical methods.

On the other hand, the quad-cascade model becomes a source of more problems when facing turbulence and other complex phenomena. For instance, in turbulent flow field with electrochemistry and electrocatalysis, the quad-cascade processes and additional characteristic length scales can make the investigations on the structures of flow and reactants more unattainable. Besides, the quad-cascade model indicates a more complex hierarchical structures relative to conventional turbulence.

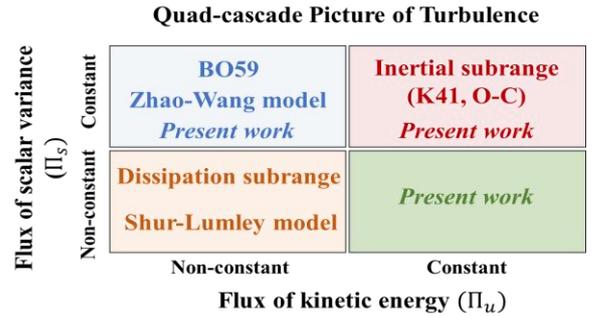

**Figure 11.** Perspective quad-cascade picture of turbulence relies on whether the fluxes of kinetic energy $\Pi_u$ and scalar variance $\Pi_\sigma$ are constant.

## 6. Conclusion

In summary, drawing inspiration from the interplay between constant and non-constant fluxes of kinetic energy and scalar variance, we have formulated a unified conservation model, demonstrating the existence of quad-cascade processes of turbulence and contributing to a comprehensive scaling framework applicable to various general turbulence scenarios. Furthermore, by investigating the electric conductivity and velocity fluctuations, we show the experimental evidence of quad-cascade processes of turbulence regarding both constant and non-constant $\Pi_u$ and $\Pi_\sigma$, in a microscale EK turbulence as the model system of the coupling turbulence. Three subranges have been identified from the scalar spectrum, which are: (1) inertial subrange, (2) constant-$\Pi_\sigma$ subrange, where only the flux of scalar variance is quasi-constant, and (3) constant-$\Pi_u$ subrange, where only the flux of kinetic energy is quasi-constant. These observations complete the jigsaw of quad-cascade picture (Figure 11), which are crucial towards understanding the fundamental physics of a variety of turbulence and the development of advanced simulation methods.

EK turbulence, as a typical turbulent system where the external volume forces and velocity field are strongly coupled with a scalar field, shares similar features as buoyancy-driven turbulence, magnetohydrodynamic turbulence and etc. Thus, the experimental observations in EK turbulence can have implications for the existence of BO59 law, which still remains debating in experiments thus far. Besides, the realization of turbulence in such a simple and microscale EK flow system is not merely a novel concept in the fields of turbulence and micro/nanofluidics, it also provides an effective and high-efficiency platform with integrability, to achieve high capacity, high throughput and parallel mixing/reaction systems for chemical and biomedical engineering.

**Acknowledgement**
The investigation is supported by National Natural Science Foundation of China (No. 51927804, 12174306), the Natural Science Basic Research Program of Shaanxi (2023-JC-JQ-02).